# Coupling of Real-Time and Co-Simulation for the Evaluation of the Large Scale Integration of Electric Vehicles into Intelligent Power Systems

Felix Lehfuss, Georg Lauss, Christian Seitl, Fabian Leimgruber, Martin Nöhrer, Thomas I. Strasser
Center for Energy – Electric Energy Systems
AIT Austrian Institute of Technology, Vienna, Austria
Email: {felix.lehfuss, georg.lauss, christian.seitl, fabian.leimgruber, martin.noehrer, thomas.strasser}@ait.ac.at

*Abstract*—This paper addresses the validation of electric vehicle supply equipment by means of a real-time capable co-simulation approach. This setup implies both pure software and real-time simulation tasks with different sampling rates dependent on the type of the performed experiment. In contrast, controller and power hardware-in-the-loop simulations are methodologies which ask for real-time execution of simulation models with well-defined simulation sampling rates. Software and real-time methods are connected one to each other using an embedded software interface. It is able to process signals with different time step sizes and is called "LabLink". Its design implies both common and specific input and output layers (middle layer), as well as a data bus (core). The LabLink enables the application of the co-simulation methodology on the proposed experimental platform targeting the testing of electric vehicle supply equipment. The test setup architecture and representative examples for the implemented co-simulation are presented in this paper. As such, a validation of the usability of this testing platform can be highlighted aiming to support a higher penetration of electric vehicles.

## I. Introduction

The mobility system of the future will have very strong synergies with the electric energy system [1]. It is expected that electric mobility will be one major part of tomorrows mobility system. Keeping the 2 °C temperature climate goal in mind it is expected that by 2030 in Europe 60% of all sold vehicles have a battery that can be charged via the electric power grid [2]. This high penetration level includes all vehicles that have the ability to be charged with electricity from the power grid, namely Full Battery Vehicles (BEV) and Plug in Hybrid Vehicles (PHEV). The expected rise in the penetration of electric mobility generates near future challenges for todays power distribution grids. High power Electric Vehicle Supply Equipment (EVSE) is expected to be mainly connected directly to the Medium Voltage (MV) distribution grid whereas low power private or semi-private charging stations will be connected directly to the Low Voltage (LV) distribution grid. Previous research and demonstration projects have shown that the rise in Electric Vehicles (EV) penetration will cause much more of a power issue than an energy issue (e.g., the Danish Nikola project [3], the Austrian V2G-Strategies project [4], or the European PlanGridEV [5]). These issues have the potential to affect the aforementioned distribution grids massively.

In order to mitigate the cost expensive "coper based" expansion of power distribution grids approaches like smart charging and Vehicle to Grid (V2G) are often discussed who base on the concept of externally controlling the charging process of the EV [6]. Both, smart charging as well as V2G, are very communication intense and prone to interoperability issues. These issues are tackled by high level communication standards like IEC 15118 and IEC 61850 [7].

The electric mobility environment at its expected merge with the smart grids presents a very novel problem for which solutions need to be derived and tested [8]. Regarding the testing of large scale applications co-simulation presents a method that has proven to be a valid approach [9], [10]. Never the less large co-simulations are prone to model uncertainties and therefore often lack accuracy or trustworthiness of results, as small errors in single models can have a large impact on the total simulation result [11].

On the other hand real-time Hardware-in-the-Loop (HIL) validation approaches [12], [13] present a methodology were the accuracy of results is very high but the scalability of the test environment is limited to a certain extend. In this contribution an advanced approach of combining these two simulation methods is conducted. The main benefits of such a combination of co-simulation and HIL are discussed as well as the expected drawbacks. A first implementation of this approach is presented which focuses on the grid integration of EVs with high penetration scenarios.

Following this introduction, an overview of the architecture and setup of typical real-time simulation and HIL-based methodologies applied to the domain of power systems is given in Section II. It discusses applicable software and real-time simulation methods followed by Section III with an introduction of the proposed embedded software interface called "LabLink" which is able to couple various computational tasks with different sampling times. In Section IV, an experimental application for validating the proposed co-simulation methodology is presented. It consists of various models such as large LV power distribution grids, grid operators control, or electric vehicle simulations which are executed independently; however, they are coupled via the LabLink message bus system. Finally, the paper is concluded with a summary of the main findings in Section V.

## II. REAL-TIME SIMULATION AND HIL APPROACHES

In general, real-time computing platforms are widely used for manifold simulation techniques. Co-simulation and real-time HIL-based experiments are methods which gain increasing importance in rapid prototyping, multi-components information transfer emulation, and system pre-compliance testing [12]–[14]. The following methodologies and approaches can be realized for the validation of control programs using simulation methods [12], [13], [15], [16]:

- *Controller Hardware-in-the-Loop (CHIL):* The simulation of the power system is usually carried out on a PC-based hardware or on a so-called Digital Real-Time Simulator (DRTS), while the corresponding control solutions and algorithms are executed on the targeted (embedded) controller platform.
- *Software-in-the-Loop (SIL):* The power system and the controller simulation are executed on the same PC or DRTS hardware.
- *Co-simulation:* The power system and the controller simulation can be distributed on different PC-based or DRTS hardware simulation environments.

The CHIL simulation approach is a real-time capable methodology mainly used for rapid prototyping and the specific verification of digital controllers [12], [13], [15]. It combines many advantages of hardware testing methods and software simulation methods. The basic idea is comprised that a specific control board runs on its target platform, while it is tested versus a simulated model of the embedded system.

Another sophisticated testing and validation approach includes also real power hardware, called Power Hardware-in-the-Loop (PHIL) [14], [17]. It mainly consists of the integration of a power interface which includes an amplification unit and a signal measurement in order to connect real power equippment (EVSE, distributed generation units, loads, storages, etc.) to the overall setup. However, this structural differential implies a huge impact from a system theory point of view. The main differences between both real-time capable simulation approaches are summarized in Fig. 1.

As previously outlined, a PHIL simulation system consists of multiple blocks representing structural areas connected one to each other via interacting signals. Hereby, three major parts are characteristic for PHIL experiments which are [14], [17]:

- *Real-time simulation models:* The major component of the software side is constituted by the DRTS calculating simulated models in real-time. Its task is clearly defined as it must maintain the guaranteed real-time capability at all times, collect incoming signals, and process outgoing signals in the given time step. The link to the power interface consists of I/O ports transferring in and outgoing signals for PHIL simulation.
- *Power Interface (PI):* The task of the PI is comprised in the bidirectional, self-contained link from the software to the hardware part, and the suitable, purposive Interface Algorithm (IA) that is inherently interlinked to the Power Amplifier (PA). The key component of the PI is the PA featuring specific signal bandwidths, delay times, power and protection limits. The PI contains both forward and backward signal paths with I/O ports, conditioning circuits, and signal measurement devices.
- *Power hardware:* This part can consist of simple passive components, standard testing equipment, or custom devices. It can be assembled out of various passive or active components for particular subsystems or even out of entire systems. It represents the target of the simulation and has to exist in physical hardware being inherently connected to the PI. The link to the PI is stated via dedicated signal measurement paths in the feedback loop and the power connection in the feed-forward loop.

## III. REAL-TIME CO-SIMULATION SYSTEM

In this paper, a coupling of the above-mentioned software and real-time simulation methods is applied to EVSE (i.e., charging platforms) in order to evaluate the expected large-scale roll-out of EVs. The general architecture and setup with the main elements is outlined in Fig. 2. Software simulations with different tasks having corresponding sampling rates of $t_{S,O1} \ldots t_{S,O(N-1)}$, while real-time simulation tasks have a defined rates of $t_{S,RT}$. The connection between pure software and real-time simulation tasks is effectuated via the so-called "LabLink" approach representing an interface layer which allows the asynchronous coupling of various simulation tools and environments running with different sampling rates.

Fig. 3 details the proposed architecture of the LabLink which is designed as embedded interfacing software connecting each software simulation task with the real-time domain. Its design includes an input interface (middle layer) with linked customized input blocks (i.e., *in, 1 . . . in, N*) feeding the data bus (core). The output interface (middle layer) equally communicates with output blocks customized for interaction with real-time simulations (i.e., *out, 1 . . . in, N*).

An overview of the integrated simulation tools and laboratory hardware at the AIT SmartEST lab is outlined in Fig. 4. The top section of the image shows the hardware part of the laboratory consisting power components that allow to interface with the electric energy path (i.e., grid emulator, charging unit, EV). In the lower part of the figure the simulation environment for analysis and evaluating EV charging processes – with respect to the grid behaviour – is shown (i.e., electric grid, EVSE operation, and EV behaviour simulation). In addition, a Supervisory Control and Data Acquisition (SCADA) system is used to control all hardware components within the laboratory and it provides a common link to the simulation environment. All components of the EV charging system within the laboratory environment exist as real-world components, as emulators, or can be executed in pure software simulation.

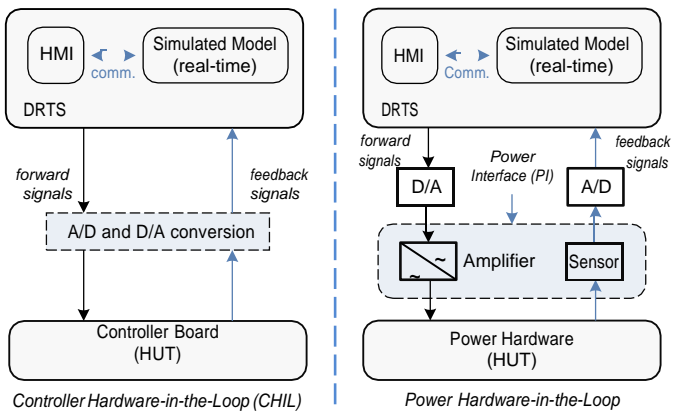

Fig. 1. Structure of real-time HIL-based simulation concepts: CHIL (left) vs. PHIL (right) [12], [17].

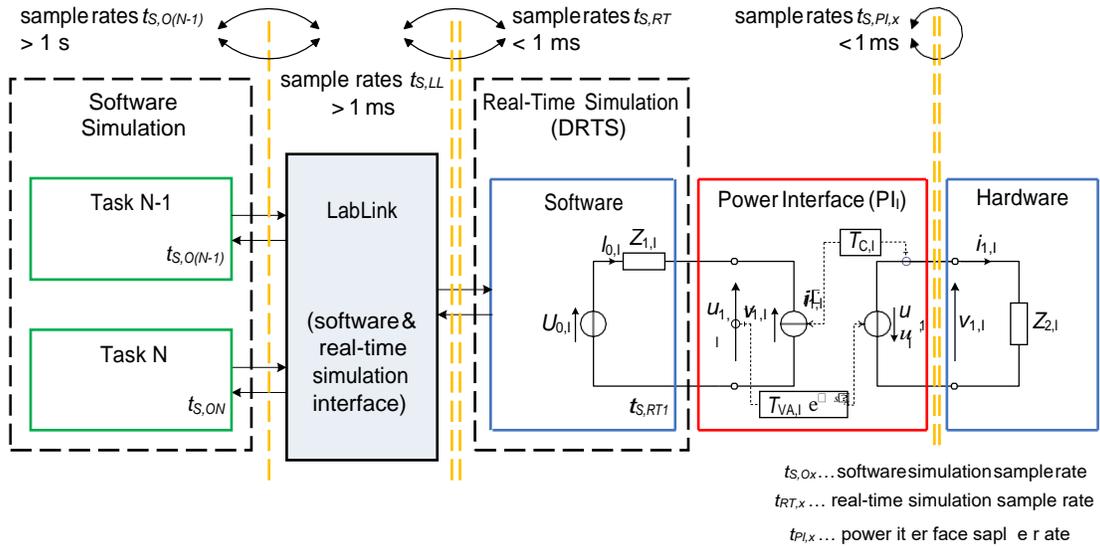

Fig. 2. Concept of a real-time PHIL simulation platform for EVSE.

For this work, the existing laboratory and simulation environment is decoupled at the SCADA part. For the purpose of decoupling, the connection of the interface signals between the simulation part implemented in software and the physical hardware part is then realized with a real-time simulation system. In general, the proposed design features a highly modular approach used for various validation and test cases like interoperability testing, charge cycle analysis, and a virtual connection to EV and/or smart grid clouds.

## IV. PROOF-OF-CONCEPT APPLICATION EXAMPLE

The experimental application (as depicted in the following Fig. 5) includes a LV power distribution grid, a grid operator, an EVSE and EV simulation as well as an EVSE operator in an simulation environment (non real-time). Each of these models and elements are executed separately in their corresponding, own simulation environments and tools. The coupling of them is realized using the above introduced LabLink message bus system. In addition to these large scale simulations a single EV and EVSE unit (i.e., charging station) are connected to a power amplifier as real devices according to Fig. 4. The power amplification unit receives input signals that represent the voltage at a selected grid node. These input signals are routed via the LabLink system into a DRTS which then computes the control signal for the power amplification.

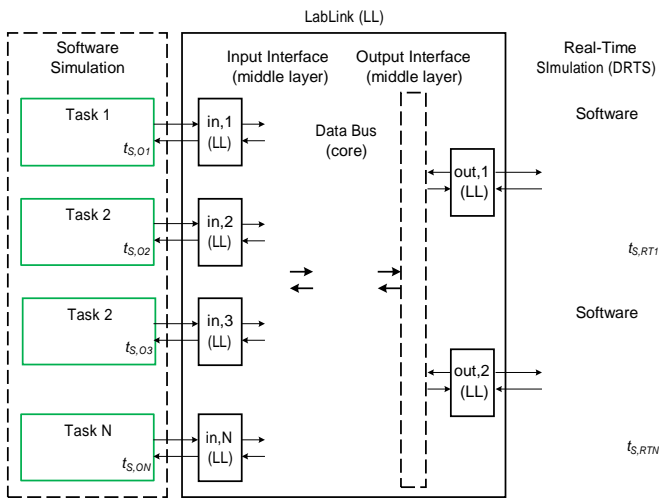

Fig. 3. Architecture of the LabLink implementation for validating real-time applications in power systems.

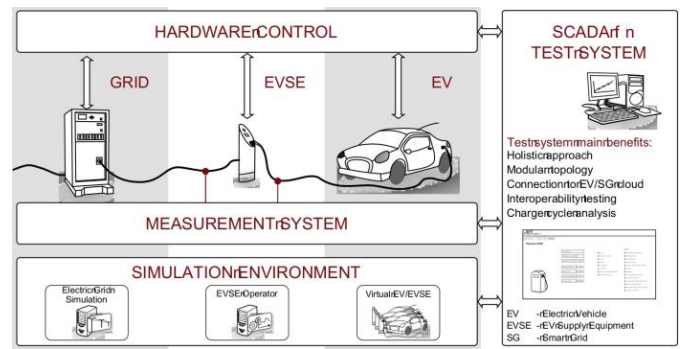

Fig. 4. Architectural overview of the utilized simulation and laboratory environment.

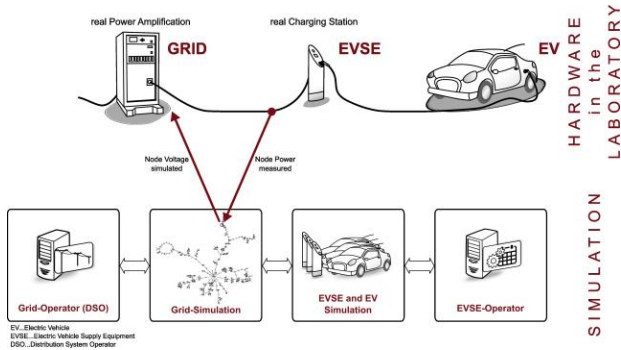

Fig. 5. Schematic overview of the implemented test-case example.

This exemplary test case enables to combine the ability to evaluate the interaction of large amounts of EVs charging in LV power distribution grids. This type of scenario can be utilized to test different aspects like:

- The feasibility and implement ability of different communication protocols and methods.
- The validation of different smart charging methods in order to minimize the effect of large EV penetration to LV power distribution grids (or to keep the effects in manageable amplitudes).
- Increase the accuracy of simulation models with the use of real components.
- Evaluate that real components who have certain communication and power behavior are really able to interact with large systems.

For this work a simplified proof-of-concept implementation is realized. It uses a small LV distribution grid as depicted in Fig. 6. It consists out of two small feeders with four loads in total of which three have a charging station. CS1 and SC2 are connected to Node 1 and CS3 is connected to Node 2. Distribution lines 1 and 2 implemented in this setup have deliberately be chosen to be extensible long such that the charging of a single vehicle already significantly affects the node voltage such that the implemented LV grid already has voltage issues with only three EVs charging at the same time.

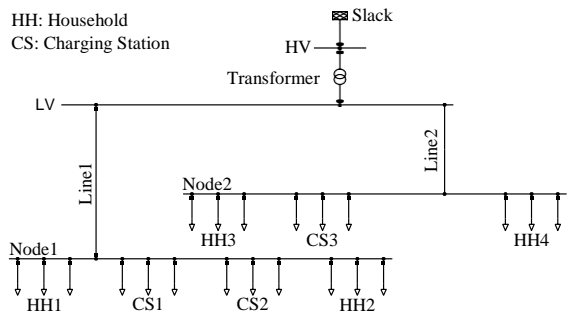

Fig. 6. Depiction of the low voltage grid implemented for the poof of concept implementation.

This scenario is executed in three different set-ups:
- As a HIL set-up with a simulated power distribution grid and a real EV and EVSE unit.
- As a HIL set-up with a simulated power distribution grid, an emulated EV, and a real EVSE unit.
- As a pure software simulation.

As two out of these three scenarios need to be executed in wall clock time it was chosen to concentration the evaluation to the time frame of the late afternoon from 5:30 pm until 9 pm. This would resemble average users that return home after work and initiate the charging process of their EV.

The execution of the same scenario by these three setups – as described above – allows for a direct comparison of the results using a real EV and EVSE unit, emulations of them, or just pure software simulations. This comparison is an important step towards the scale-up of such scenarios. A large scale implementation of the proposed set-up, but without a HIL coupling can be found for example in [18].

Fig. 7 and Fig. 8 show the achieved results of these three scenarios. For this demonstration level implementation no charge control algorithm was implemented, thus the results show the effects of uncontrolled charging onto the implemented LV distribution grid. Fig. 7 shows the effects that the EV charging has on the phase 1 voltage of node 1 when two EV charge at CS1 and CS2. The CS1 was the charging station that was connected to the power amplification unit.

Evaluating the charging power (see Fig. 8) a difference of 10% is visible between the real EV and the simulated one as well as between the real EV and the emulated one.

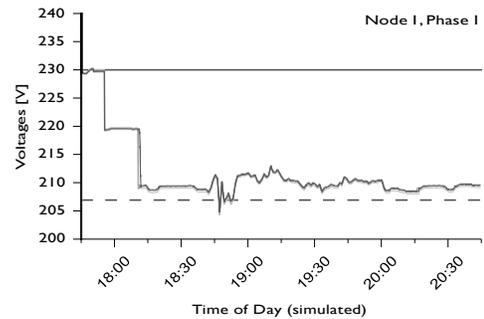

Fig. 7. Resulting voltage at node 1 of the proof of concept implementation.

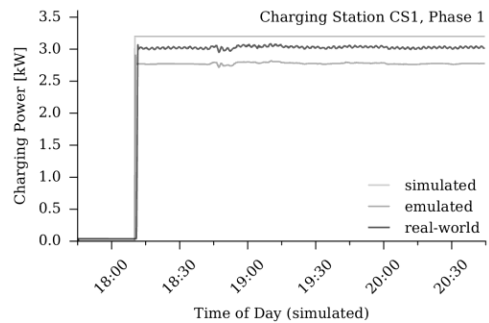

Fig. 8. Resulting charging power of the proof of concept implementation.

Although the error could be easily corrected within the model parameterization as it is a simple offset type of error. However, one of the main advantages of this co-simulation/HIL coupled setup is that such kind of errors cannot occur if the utilized EV is a real one.

Summarizing, the proposed configuration in this work allows for sophisticated in depth analysis of the effects of different charge control algorithms for the large scale grid integration of electric mobility. One of the key assets is the fact that for such a large scale implementation single vehicles can be connected as real hardware resulting in a kind of online validation tool. The proposed setup is not limited to be used for grid integration of electric mobility but it can also be used for various other purposes where the transient effects of multiple loads are triggered simultaneously as for example:

- Evaluation of multiple home management systems in a local neighborhood.
- Evaluation of different flexibility market products and their transient effects (especially when price signals trigger different loads or generators).
- Evaluation of smart grid technologies and control solutions on LV power distribution grid levels.

## V. Summary and Conclusions

This paper presents a co-simulation approach for the validation of EVSE and corresponding (smart) charging concepts. Different simulation and HIL-based methodologies are itemized and explained. As a first result, the so-called LabLink is introduced as an embedded software interface. It shows the capability to link all independently executed calculation tasks one to each other resulting in valid requirements for a real-time capable co-simulation platform. The architecture of the LabLink is highlighted and different layers are detailed for the input and output sides.

As a second main contribution of this work, the implementation of a testing platform for the ecosystem of electric mobility and its interaction with the electric power grids is proposed.

The novel integration of a real-time into a co-simulation environment enables a novel testing procedure for the holistic evaluation of EV grid integration. In this work only a small scale example is given, but due to the architecture of the proposed system the limitation in the system-scale in only given by the computational power of the single device that is utilized to simulate the single simulation as different devices/computers/simulation tools are coupled via the LabLink framework. Coupling a DRTS to this framework allows a highly accurate and fully transient implementation of single devices to such a large scale simulation and thereby allows to test the effects of large scale control onto single devices or vice-versa the effects of single device controls or behavior on large electric energy infrastructure.

## Acknowledgment

This work is supported by the European Communitys Horizon 2020 Program (H2020/2014-2020) under project "ERI-Grid" (Grant Agreement No. 654113).